\pdfoutput=1

\documentclass[12pt,a4paper]{article}

\usepackage{ifthen} 
\newboolean{pdflatex}
\setboolean{pdflatex}{true} 
\usepackage{graphicx}
\usepackage{epstopdf}
\usepackage{multirow}
\usepackage{rotating}
\usepackage{bigints}

\newboolean{articletitles} 
\setboolean{articletitles}{true} 

\newboolean{uprightparticles}
\setboolean{uprightparticles}{false} 

\def\paperauthors{LHCb collaboration} 
\def\papercopyright{\the\year\ CERN for the benefit of the LHCb collaboration} 
\def\paperlicence{CC-BY-4.0 licence}
\def\paperlicenceurl{https://creativecommons.org/licenses/by/4.0/}

\usepackage{microtype}
\usepackage{lineno}  
\usepackage{xspace} 
\usepackage{caption} 

\usepackage{graphicx}  
\usepackage{color}
\usepackage{colortbl}
\graphicspath{{./figs/}} 

\usepackage{amsmath} 
\usepackage{amssymb}
\usepackage{amsfonts}
\usepackage{upgreek} 

\newcommand*\patchAmsMathEnvironmentForLineno[1]{%
  \expandafter\let\csname old#1\expandafter\endcsname\csname #1\endcsname
  \expandafter\let\csname oldend#1\expandafter\endcsname\csname
  end#1\endcsname
  \renewenvironment{#1}%
  {\linenomath\csname old#1\endcsname}%
  {\csname oldend#1\endcsname\endlinenomath}%
}
\newcommand*\patchBothAmsMathEnvironmentsForLineno[1]{%
  \patchAmsMathEnvironmentForLineno{#1}%
  \patchAmsMathEnvironmentForLineno{#1*}%
}
\AtBeginDocument{%
  \patchBothAmsMathEnvironmentsForLineno{equation}%
  \patchBothAmsMathEnvironmentsForLineno{align}%
  \patchBothAmsMathEnvironmentsForLineno{flalign}%
  \patchBothAmsMathEnvironmentsForLineno{alignat}%
  \patchBothAmsMathEnvironmentsForLineno{gather}%
  \patchBothAmsMathEnvironmentsForLineno{multline}%
  \patchBothAmsMathEnvironmentsForLineno{eqnarray}%
}

\usepackage{hyperref}    
\usepackage[all]{hypcap} 
\usepackage{bold-extra}  
\usepackage{xspace} 
\usepackage{upgreek}

\def\lhcb   {\mbox{LHCb}\xspace}

\def\velo   {VELO\xspace}

\def\MagUp {\mbox{\em Mag\kern -0.05em Up}\xspace}

\def\hltone {HLT1\xspace}

\ifthenelse{\boolean{uprightparticles}}%
{

 \def\Pmu         {\ensuremath{\upmu}\xspace}

 \def\Ppi         {\ensuremath{\uppi}\xspace}

 \def\Ppsi        {\ensuremath{\uppsi}\xspace}

 \def\PDelta      {\ensuremath{\Delta}\xspace}                 
 \def\PXi         {\ensuremath{\Xi}\xspace}                 
 \def\PLambda     {\ensuremath{\Lambda}\xspace}                 
 \def\PSigma      {\ensuremath{\Sigma}\xspace}                 
 \def\POmega      {\ensuremath{\Omega}\xspace}                 
 \def\PUpsilon    {\ensuremath{\Upsilon}\xspace}
 \let\oldPi\Pi
 \def\PPi         {\ensuremath{\oldPi}\xspace}

 \def\PB      {\ensuremath{\mathrm{B}}\xspace}                 
                  
 \def\PD      {\ensuremath{\mathrm{D}}\xspace}

 \def\PJ      {\ensuremath{\mathrm{J}}\xspace}                 
 \def\PK      {\ensuremath{\mathrm{K}}\xspace}

 \def\Pb      {\ensuremath{\mathrm{b}}\xspace}                 
 \def\Pc      {\ensuremath{\mathrm{c}}\xspace}                 
                  
 \def\Pe      {\ensuremath{\mathrm{e}}\xspace}

 \def\Pi      {\ensuremath{\mathrm{i}}\xspace}

 \def\Ps      {\ensuremath{\mathrm{s}}\xspace}

 \def\thebaroffset{0.0em}
}
{

 \def\Pmu         {\ensuremath{\mu}\xspace}

 \def\Ppi         {\ensuremath{\pi}\xspace}

 \def\Ppsi        {\ensuremath{\psi}\xspace}                 
                  
 \mathchardef\PDelta="7101
 \mathchardef\PXi="7104
 \mathchardef\PLambda="7103
 \mathchardef\PSigma="7106
 \mathchardef\POmega="710A
 \mathchardef\PUpsilon="7107
 \mathchardef\PPi="7105
                  
 \def\PB      {\ensuremath{B}\xspace}                 
                  
 \def\PD      {\ensuremath{D}\xspace}

 \def\PJ      {\ensuremath{J}\xspace}                 
 \def\PK      {\ensuremath{K}\xspace}

 \def\Pb      {\ensuremath{b}\xspace}                 
 \def\Pc      {\ensuremath{c}\xspace}                 
                  
 \def\Pe      {\ensuremath{e}\xspace}

 \def\Pi      {\ensuremath{i}\xspace}

 \def\Ps      {\ensuremath{s}\xspace}

 \def\thebaroffset{0.18em}
}
\newcommand{\offsetoverline}[2][\thebaroffset]{\kern #1\overline{\kern -#1 #2}}%

\makeatletter
\ifcase \@ptsize \relax
  \newcommand{\miniscule}{\@setfontsize\miniscule{4}{5}}
\or
  \newcommand{\miniscule}{\@setfontsize\miniscule{5}{6}}
\or
  \newcommand{\miniscule}{\@setfontsize\miniscule{5}{6}}
\fi
\makeatother

\DeclareRobustCommand{\optbar}[1]{\shortstack{{\miniscule (\rule[.5ex]{1.25em}{.18mm})}
  \\ [-.7ex] $#1$}}

\def\en         {{\ensuremath{\Pe^-}}\xspace}   
\def\ep         {{\ensuremath{\Pe^+}}\xspace}

\def\mup        {{\ensuremath{\Pmu^+}}\xspace}
\def\mun        {{\ensuremath{\Pmu^-}}\xspace} 
 
\def\mump       {{\ensuremath{\Pmu^\mp}}\xspace}

\def\squark    {{\ensuremath{\Ps}}\xspace}

\def\cquark    {{\ensuremath{\Pc}}\xspace}

\def\bquark    {{\ensuremath{\Pb}}\xspace}

\def\pion   {{\ensuremath{\Ppi}}\xspace}

\def\pip    {{\ensuremath{\pion^+}}\xspace}
\def\pim    {{\ensuremath{\pion^-}}\xspace}
\def\pipm   {{\ensuremath{\pion^\pm}}\xspace}

\def\kaon    {{\ensuremath{\PK}}\xspace}

\def\KorKbar {\kern \thebaroffset\optbar{\kern -\thebaroffset \PK}{}\xspace}

\def\KS      {{\ensuremath{\kaon^0_{\mathrm{S}}}}\xspace}

\def\KL      {{\ensuremath{\kaon^0_{\mathrm{L}}}}\xspace}

\def\D       {{\ensuremath{\PD}}\xspace}

\def\DorDbar {\kern \thebaroffset\optbar{\kern -\thebaroffset \PD}\xspace}

\def\Dp      {{\ensuremath{\D^+}}\xspace}
\def\Dm      {{\ensuremath{\D^-}}\xspace}

\def\DpDm    {\ensuremath{\Dp {\kern -0.16em \Dm}}\xspace}

\def\B       {{\ensuremath{\PB}}\xspace}

\def\BorBbar {\kern \thebaroffset\optbar{\kern -\thebaroffset \PB}\xspace}

\def\Bd      {{\ensuremath{\B^0}}\xspace}

\def\BdorBdbar {\kern \thebaroffset\optbar{\kern -\thebaroffset \Bd}\xspace}

\def\Bs      {{\ensuremath{\B^0_\squark}}\xspace}

\def\BsorBsbar {\kern \thebaroffset\optbar{\kern -\thebaroffset \Bs}\xspace}

\def\jpsi     {{\ensuremath{{\PJ\mskip -3mu/\mskip -2mu\Ppsi}}}\xspace}

\def\Y#1S{\ensuremath{\PUpsilon{(#1S)}}\xspace}

\def\Lz          {{\ensuremath{\PLambda}}\xspace}

\def\LorLbar     {\kern \thebaroffset\optbar{\kern -\thebaroffset \PLambda}\xspace}

\def\to                 {\ensuremath{\rightarrow}\xspace}

\def\AT#1     {\ensuremath{A_{\mathrm{T}}^{#1}}\xspace}           

\def\C#1      {\ensuremath{\mathcal{C}_{#1}}\xspace}                       
\def\Cp#1     {\ensuremath{\mathcal{C}_{#1}^{'}}\xspace}                    
\def\Ceff#1   {\ensuremath{\mathcal{C}_{#1}^{\mathrm{(eff)}}}\xspace}        
\def\Cpeff#1  {\ensuremath{\mathcal{C}_{#1}^{'\mathrm{(eff)}}}\xspace}       
\def\Ope#1    {\ensuremath{\mathcal{O}_{#1}}\xspace}                       
\def\Opep#1   {\ensuremath{\mathcal{O}_{#1}^{'}}\xspace}                    

\newcommand{\nospaceunit}[1]{\ensuremath{\text{#1}}}       
\newcommand{\aunit}[1]{\ensuremath{\text{\,#1}}}       

\newcommand{\tev}{\aunit{Te\kern -0.1em V}\xspace}
\newcommand{\gev}{\aunit{Ge\kern -0.1em V}\xspace}
\newcommand{\mev}{\aunit{Me\kern -0.1em V}\xspace}
\newcommand{\kev}{\aunit{ke\kern -0.1em V}\xspace}
\newcommand{\ev}{\aunit{e\kern -0.1em V}\xspace}
 
\newcommand{\mevc}{\ensuremath{\aunit{Me\kern -0.1em V\!/}c}\xspace}
\newcommand{\gevc}{\ensuremath{\aunit{Ge\kern -0.1em V\!/}c}\xspace}
\newcommand{\mevcc}{\ensuremath{\aunit{Me\kern -0.1em V\!/}c^2}\xspace}
\newcommand{\gevcc}{\ensuremath{\aunit{Ge\kern -0.1em V\!/}c^2}\xspace}

\def\mum  {\ensuremath{\,\upmu\nospaceunit{m}}\xspace}

\def\fb   {\ensuremath{\aunit{fb}}\xspace}
\def\invfb   {\ensuremath{\fb^{-1}}\xspace}

\newcommand{\chisq}{\ensuremath{\chi^2}\xspace}

\newcommand{\chisqip}{\ensuremath{\chi^2_{\text{IP}}}\xspace}

\def\gsim{{~\raise.15em\hbox{$>$}\kern-.85em
          \lower.35em\hbox{$\sim$}~}\xspace}
\def\lsim{{~\raise.15em\hbox{$<$}\kern-.85em
          \lower.35em\hbox{$\sim$}~}\xspace}

\def\pt         {\ensuremath{p_{\mathrm{T}}}\xspace}

\def\evtgen     {\mbox{\textsc{EvtGen}}\xspace}

\def\geant      {\mbox{\textsc{Geant4}}\xspace}

\def\photos     {\mbox{\textsc{Photos}}\xspace}

\def\pythia     {\mbox{\textsc{Pythia}}\xspace}

\def\tell1  {TELL1\xspace}
\def\ukl1   {UKL1\xspace}

\newcommand{\lhcborcid}[1]{\href{https://orcid.org/#1}{\hspace*{0.1em}\raisebox{-0.45ex}{\includegraphics[width=1em]{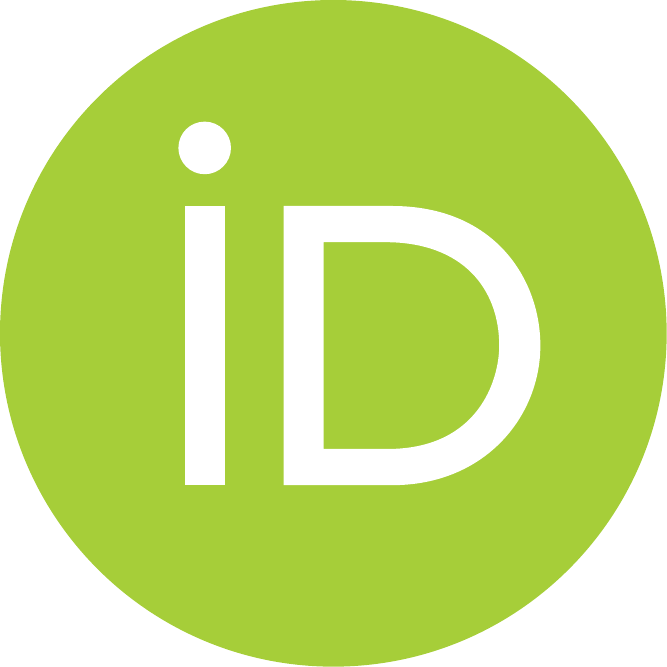}}}}

\usepackage{cite} 
\usepackage{mciteplus}

\usepackage{mathrsfs}
\usepackage{longtable} 
\usepackage{subfig}
\usepackage{booktabs}
\usepackage[rgb]{xcolor}

\newcommand{\Ks}{\KS}
\newcommand{\Kl}{\KL}

\newcommand{\Ksmumu}{\ensuremath{\Ks\to\mu^+\mu^-}\xspace}

\newcommand{\Kspipi}{\ensuremath{\Ks\to\pi^+\pi^-}\xspace}

\newcommand{\KFM}{\ensuremath{\Ks\to\mu^+\mu^-\mu^+\mu^-}\xspace}
\newcommand{\KlFM}{\ensuremath{\Kl\to\mu^+\mu^-\mu^+\mu^-}\xspace}
\newcommand{\KzFM}{\ensuremath{K^0_{\mathrm{S(L)}}\to\mu^+\mu^-\mu^+\mu^-}\xspace}

\newcommand{\KLFM}{\ensuremath{\Kl\to\mu^+\mu^-\mu^+\mu^-}\xspace}

\newcommand{\Lambdappi}{\ensuremath{\Lz\to p\pi^-}\xspace}

\newcommand{\BRof}[1]{\ensuremath{{\cal B}(#1)}\xspace}

\newcommand{\IP}{\ensuremath{{\rm IP}}\xspace}

\newcommand{\figref}[1]{Fig.~\ref{#1}}
\newcommand{\tabref}[1]{Table~\ref{#1}}

\def\nuornubar {\kern \thebaroffset\optbar{\kern -\thebaroffset \nu}\xspace} 

\begin{document}

\renewcommand{\thefootnote}{\fnsymbol{footnote}}
\setcounter{footnote}{1}


\begin{titlepage}
  \pagenumbering{roman}

  \vspace*{-1.5cm}
  \centerline{\large EUROPEAN ORGANIZATION FOR NUCLEAR RESEARCH (CERN)}
  \vspace*{1.5cm}
  \noindent
  \begin{tabular*}{\linewidth}{lc@{\extracolsep{\fill}}r@{\extracolsep{0pt}}}
    \ifthenelse{\boolean{pdflatex}}
    {\vspace*{-1.5cm}\mbox{\!\!\!\includegraphics[width=.14\textwidth]{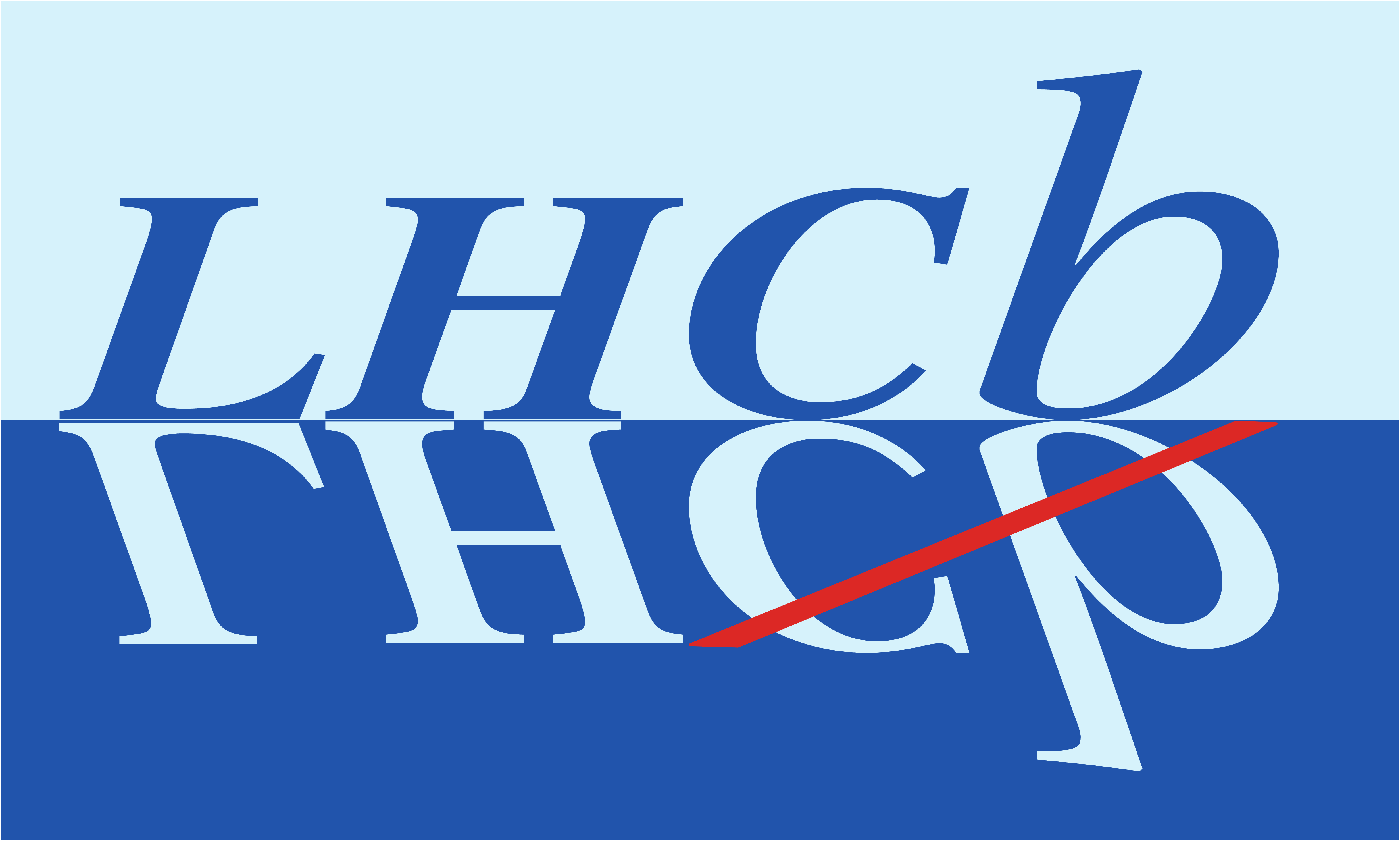}} & &}%
    {\vspace*{-1.2cm}\mbox{\!\!\!\includegraphics[width=.12\textwidth]{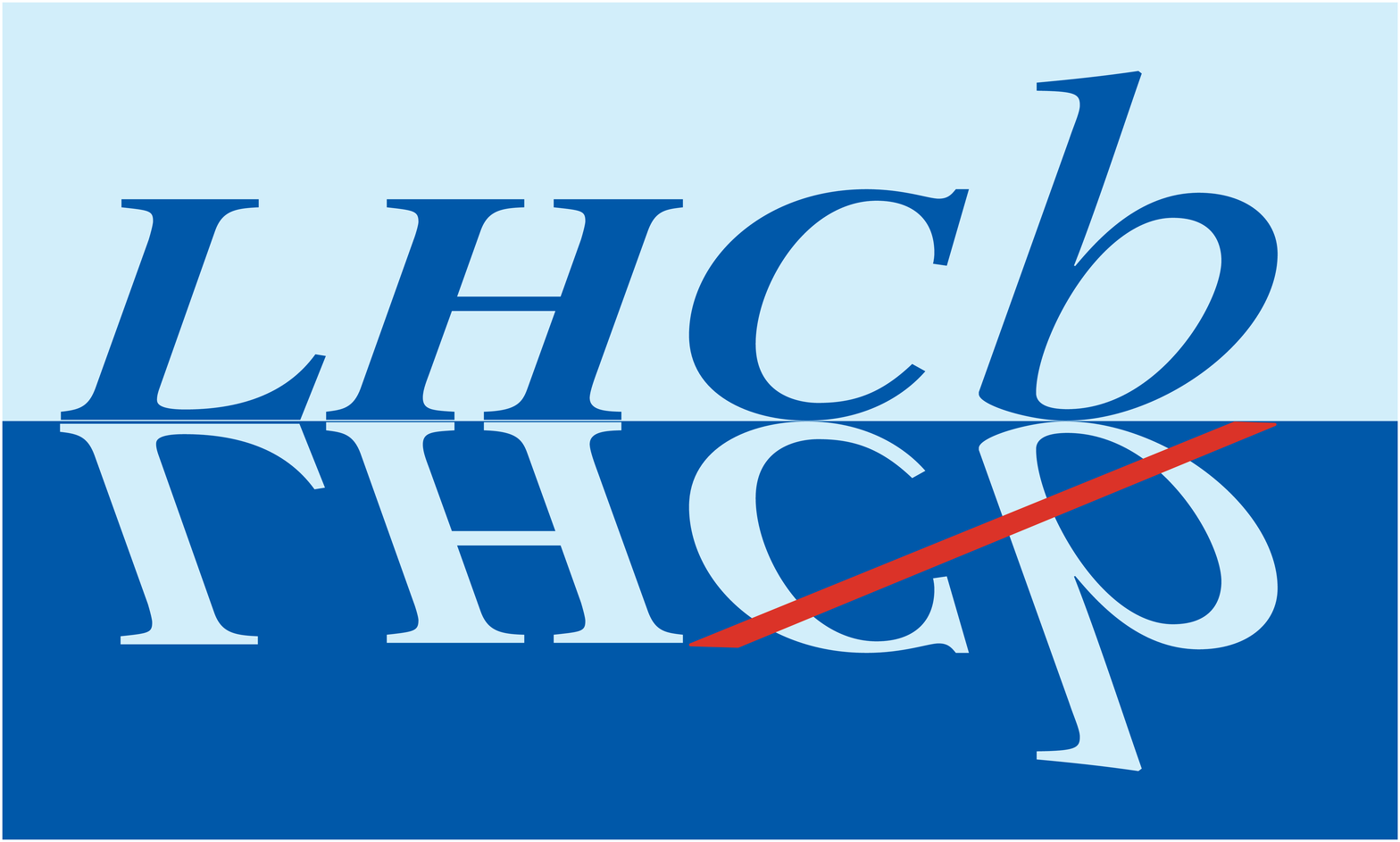}} & &}%
    \\
    & & CERN-EP-2022-255 \\  
    & & LHCb-PAPER-2022-035 \\  
    & & December 15, 2022 \\
    & & \\
  \end{tabular*}

  \vspace*{4.0cm}

  {\normalfont\bfseries\boldmath\huge
    \begin{center}
            Search for \KzFM decays at LHCb
    \end{center}
  }

  \vspace*{2.0cm}

  \begin{center}
    \paperauthors\footnote{Authors are listed at the end of this paper.}
  \end{center}

  \vspace{\fill}

  \begin{abstract}
    \noindent
    A search for $K_{\textrm{S(L)}}^0\rightarrow\mu^{+}\mu^{-}\mu^{+}\mu^{-}$ decays is performed using proton-proton collision data collected by the LHCb experiment at a centre-of-mass energy of 13 TeV, corresponding to an integrated luminosity of 5.1~\invfb. No evidence for signal is found. The 90\% confidence level upper limits are the first set for both decays and are \BRof\KFM $< 5.1\times 10^{-12}$ and \BRof\KlFM $< 2.3\times 10^{-9}$.

  \end{abstract}

  \vspace*{2.0cm}

  \begin{center}
    Published in Phys. Rev. D108 (2023) L031102
  \end{center}

  \vspace{\fill}

  {\footnotesize
    \centerline{\copyright~\papercopyright. \href{\paperlicenceurl}{\paperlicence}.}}
  \vspace*{2mm}

\end{titlepage}


\newpage
\setcounter{page}{2}
\mbox{~}
%
%
%
%

\cleardoublepage


\renewcommand{\thefootnote}{\arabic{footnote}}
\setcounter{footnote}{0}



\pagestyle{plain} 
\setcounter{page}{1}
\pagenumbering{arabic}


%

\section{Introduction}
\label{sec:Introduction}

The \KFM decay is a flavour-changing neutral current process that has not yet been observed. In the Standard Model (SM), its decay rate is highly suppressed, with an expected branching fraction~\cite{DAmbrosio:2013qmd}
\begin{equation}
  \BRof\KFM_{\text{SM}} \sim (1 \text{--} 4) \times 10^{-14}.
\end{equation}

\noindent The related channel \KLFM is predicted in the SM~\cite{DAmbrosio:2013qmd} to occur at a higher rate
\begin{equation}
  \BRof\KLFM_{\text{SM}} \sim (4 \text{--} 9)  \times 10^{-13}.
\end{equation}
\noindent However, 
since the expected LHCb acceptance for \Kl decays is $\sim 2\times10^{-3}$ times smaller than the \Ks one~\cite{Junior:2018odx},
the SM \KlFM yield expected in the experiment is small compared to that of the \Ks decay.

Physics beyond the SM (BSM) can lead to large enhancements of \BRof\KFM with respect to the SM prediction. For instance, proposed dark-sector scenarios can enhance the branching fractions by up to $\sim 2\times10^{-12}$ ~\cite{Goudzovski:2022vbt,Hostert:2020xku}. To date, no direct experimental search of these decays has been performed. This paper focuses primarily on a search for the \KFM decay, but results for both the \KFM and \KLFM decays are presented.

In this analysis, proton-proton ($pp$) collision data collected by the LHCb experiment during 2016--2018 at a centre-of-mass energy of 13\tev, corresponding to an integrated luminosity of $5.1\,\text{fb}^{-1}$, are used. The data analysis largely follows the strategy of the \mbox{\Ksmumu} analyses by the LHCb collaboration~\cite{LHCb-PAPER-2019-038, LHCb-PAPER-2012-023}. Following those, \BRof\KFM and \BRof\KlFM are measured relative to \BRof\Kspipi.

\section{LHCb detector and trigger}
The LHCb detector~\cite{LHCb-DP-2008-001,LHCb-DP-2014-002} is a single-arm forward spectrometer covering the pseudorapidity range 
$2 < \eta < 5$, designed for the study of particles containing \bquark or \cquark quarks.
The detector includes a high-precision tracking system consisting of a silicon-strip vertex detector (VELO) surrounding the $pp$ interaction region~\cite{LHCb-DP-2014-001}, a large-area silicon-strip detector located upstream 
of a dipole magnet with a bending power of about $4{\mathrm{\,Tm}}$, and three stations of silicon-strip detectors 
and straw drift tubes~\cite{LHCb-DP-2017-001} placed downstream of the magnet. The tracking system provides a measurement of the momentum, $p$, of charged particles with a relative uncertainty that varies from 0.5\% at low momentum to 1.0\% at 200~\gevc. The minimum distance of a track to a proton-proton collision vertex (PV),
the impact parameter (IP), is measured with a resolution of $(15+29/\pt)\mum$, where \pt is the component of the momentum transverse to the beam axis, in \gevc. Different types
of charged hadrons are distinguished using information from two ring-imaging Cherenkov
(RICH) detectors~\cite{LHCb-DP-2012-003}. Photons, electrons and hadrons are identified by a calorimeter system consisting of scintillating-pad and preshower detectors, an electromagnetic and a hadronic calorimeter. Muons are identified by a system composed of alternating layers of iron
and multiwire proportional chambers~\cite{LHCb-DP-2012-002}. In addition, information from the tracking system, the calorimeter system and the RICH detectors is used to further improve the muon
identification. The LHCb detector accumulated data in two data-taking periods: Run 1 (2011--2012) and Run 2 (2015--2018).

Events are first filtered by a hardware trigger, known as L0~\cite{Aaij:2012me}, which uses information from the calorimeter and the muon system to select \pt signatures above a few \gevc. Subsequently,
a full event reconstruction is applied in the High Level Trigger (HLT), which runs in two separate steps (HLT1 and HLT2). In analyses of kaon decays to muons using data recorded prior to 2016, the search was limited by a requirement of at least one muon with \pt above 1.8~\gevc at HLT1. In 2016, a new reconstruction algorithm allowed the muon \pt threshold requirement in the HLT to be reduced to 80~\mevc~\cite{Aaij:2253050}. In
addition, a dedicated set of software trigger selections was developed for the  \hltone stage~\cite{Dettori:2297352}.
This increased the trigger efficiency for selecting strange hadrons decaying into muons by an order of magnitude with respect to the LHCb Run 1 data sample~\cite{Dettori:2297352}. 

Given its large branching fraction~\cite{PDG2022}, the \Kspipi decay is used as the normalisation mode in this search. The use of this normalisation mode cancels part of the systematic uncertainties from signal detection efficiency, and does not require knowledge of the absolute kaon production cross-section and integrated luminosity.
The \Kspipi candidates are reconstructed from trigger-unbiased data selected by a quasi-random trigger with minimal requirements ensuring some event activity. The \Ks candidates from both the signal and the normalisation channel are required to come from the PV.
Simulation is used to model the effects of the detector acceptance and
selection requirements. 
In the simulation, $pp$ collisions are generated using
  \pythia8~\cite{Sjostrand:2007gs,Sjostrand:2006za} 
  with a specific \lhcb configuration~\cite{LHCb-PROC-2010-056}.
  Decays of unstable particles
  are described by \evtgen~\cite{Lange:2001uf}, in which final-state
  radiation is generated using \photos~\cite{davidson2015photos}.
  The interaction of the generated particles with the detector, and its response,
  are implemented using the \geant
  toolkit~\cite{Allison:2006ve, Agostinelli:2002hh} as described in
  Ref.~\cite{LHCb-PROC-2011-006}.

\section{Strategy}
This measurement benefits from both the large prompt \Ks and \Kl production cross-sections at the LHC (with cross-sections times multiplicity of order $0.3~\text{barn}$ each~\cite{Junior:2018odx}), as well as the forward production of kaons which fall within the LHCb detector acceptance.
Candidates of \KFM decays are reconstructed from two pairs of muons with opposite charges, forming a sufficiently detached secondary vertex with an invariant mass lower than 600\mevcc.
A blinding procedure is followed: selected candidates in the four-muon invariant-mass region $490 < m_{4\mu} < 510\mevcc$, which, estimated from simulation, contains $\approx 97\%$ of the signal,
are removed from the data sample until the analysis procedure is finalized. The four-muon mass resolution is approximately 4\mevcc.

The data sample is split according to the L0 hardware trigger decision: a category where an object in the event other than the muons from the \KFM candidate fires L0 (triggered independently of signal, TIS), and a category where the muons from \KFM candidates satisfy muon-related L0 requirements (exclusively triggered on signal, xTOS)~\cite{LHCb-2008-073}.
At the HLT stage, the trigger decision must be caused by one of the signal muon candidates. The analysis is performed separately on two independent datasets, labelled TIS and xTOS in reference to the L0 decision, hereafter referred to as trigger categories. Each category comprises roughly 50\% of the signal candidates. Since decays occurring inside the VELO can be reconstructed with better invariant-mass resolution than decays outside the VELO, and because the trigger relies on tracks with hits within the VELO and downstream tracking stations, only decays occurring inside the VELO are considered in this analysis.
The LHCb decay-time acceptance is such that it makes \KlFM and \KFM decays difficult to differentiate~\cite{Junior:2018odx}. For this reason, \KlFM decays are neglected when setting an upper limit on \BRof\KFM, and vice versa. This approach yields conservative upper limits in the presence of signal. If a signal were to be observed, it would be identified as \KzFM, without specifying the $K^0$ mass eigenstate. 

\section{Signal selection}
\label{sec:selection}
Several selection requirements are applied to the signal candidate events to further reduce background contributions. These include requirements on the track quality, track \chisqip (defined as the difference in the vertex-fit \chisq of a given PV reconstructed with and without the track being considered), and particle identification (PID) of the muons. The \Kspipi candidates are also selected using track quality, \chisqip, and PID requirements, in addition to a requirement on the variables from the Armenteros-Podolanski plane~\cite{armenteros} to reduce contributions from \Lambdappi decays.

The dominant background source in the signal channel arises from random combinations of tracks originating near a $pp$ interaction region or from inelastic interactions with the detector material.
Potential peaking background contributions from $K^{0}\rightarrow\pip\pim\mup\mun$, $K^{0}\rightarrow\mup\mun \ep\en$ and $K^{0}\rightarrow\pip\pim\ep\en$ decays are studied using simulation and are found to be negligible.
A boosted decision tree~(BDT)~\cite{Breiman,AdaBoost} classifier implemented in the TMVA
toolkit~\cite{Hocker:2007ht,TMVA4} is used to further separate signal from combinatorial background. One BDT per trigger category (TIS and xTOS) is trained. The discriminating variables included in the BDTs are:

\begin{itemize}
  \item \label{var:vmv} the significance of the distance from the candidate decay vertex to any VELO element or the RF-foil~\cite{Alexander:2018png}, to eliminate background from material interactions;
  \item the smallest \IP of each of the four muon candidates with respect to any of the PVs reconstructed in the event, to eliminate background from final state tracks originating close to the $pp$ interaction point;
  \item the smallest \IP of the \Ks candidates to any of the PVs reconstructed in the event, to discriminate against background that does not point to a primary vertex;
  \item the transverse displacement of the \Ks candidate decay vertex from the beam line, to eliminate background from tracks close to the $pp$ interaction point;
  \item the maximum distance of closest approach between the four final-state tracks, which is expected to be larger for background from random combinations of tracks;
  \item and the minimum angle between each pair of muons, which significantly reduces background contributions with at least two \velo tracks sharing hits.
\end{itemize}

The BDT is trained with simulated \KFM decays as a proxy for signal, and \KFM candidates from the invariant-mass sidebands in data, which cover $450 < m_{4\mu} < 490\mevcc$ and $510 < m_{4\mu} < 600\mevcc$, as a proxy for background.
The samples are randomly split into two subsamples for cross-validation~\cite{scikit-learn}, to ensure that the BDT applied to a candidate is always trained on events that are independent of the candidate.

The BDT requirement is optimized for the best expected limit at a 90\% confidence level (C.L.) using simulated pseudoexperiments. The signal efficiency of the BDT requirement is approximately $80\%$ while retaining less than one per mille of the background candidates. After the BDT selection requirement, in the case of multiple candidates per event, a single candidate, randomly chosen, is retained.

\section{Normalisation and invariant-mass fit}
\label{sec:normalisation}
The signal yield, $N_{\KFM}$, is translated to a branching fraction via a normalisation to \Kspipi decays, as
\begin{equation}
    \BRof\KFM \equiv \alpha N_{\KFM}.
\end{equation}
\noindent The normalisation factor $\alpha$, or single event sensitivity,\footnote{The value of \BRof\KFM corresponding to one signal event seen in the data sample.}  is calculated as
\begin{equation}
    \alpha = \BRof\Kspipi\times\frac{\varepsilon_{\Kspipi}\times s_{\text{MB}}}{N_{\Kspipi}^{\text{MB}}\times{\varepsilon_{\KFM}}},
    \label{eq:alpha}
\end{equation}
\noindent where $\varepsilon_{\Kspipi} (\varepsilon_{\KFM})$ represents the product of the reconstruction, trigger and selection 
efficiency of \Kspipi (\KFM) candidates. The observed number of \Kspipi decays is $N_{\Kspipi}^{\text{MB}}$, \BRof\Kspipi is the branching fraction of the normalisation channel, fixed to $(69.20\pm0.05)\%$~\cite{PDG2022}, and $s_{\text{MB}}$ is the prescale factor of the minimum bias trigger averaged over the different data files weighted by integrated luminosity.

\begin{figure}
    \centering
    \includegraphics[width=0.7\textwidth]{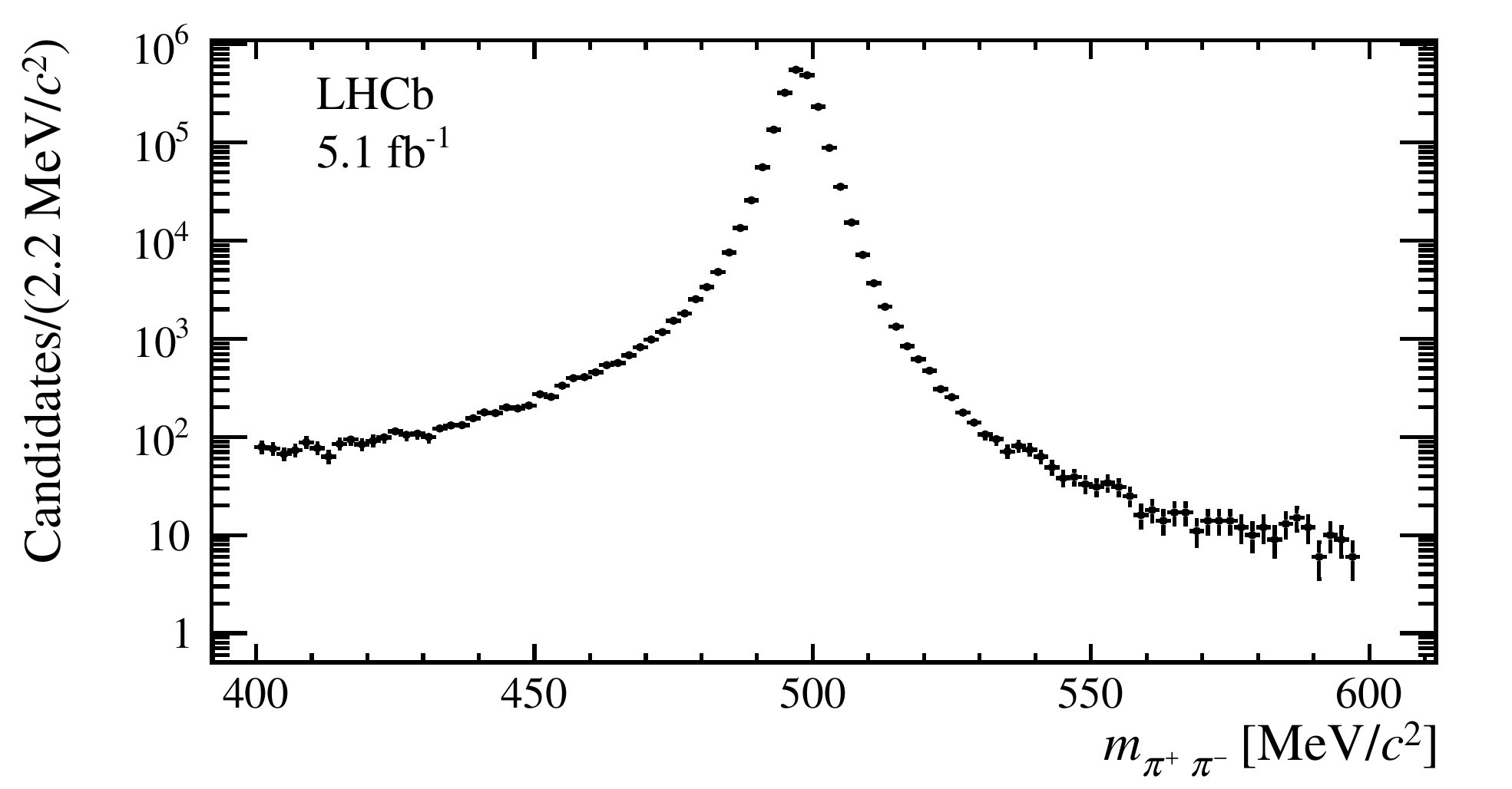}
   
    \caption{Invariant-mass distribution for the \Kspipi selected candidates in the full dataset. The background level is estimated to be at the $1\%$ level or below.}
    \label{fig:MB_Yields}
\end{figure}

The observed number of \Kspipi decays, $N_{\Kspipi}^{\text{MB}}$, is estimated as the number of candidates in the mass range $400<m_{\pip\pim}<600\mevcc$, without a fit to the invariant-mass distribution. The background is neglected due to the high purity obtained after applying the same track quality requirements as in the signal channel, as well as requirements on the transverse displacement of the \KS decay vertex from the beam line and the distance of the \Ks decay vertex from any detector material. Figure~\ref{fig:MB_Yields} shows the invariant-mass distribution of the \Kspipi candidates satisfying these requirements.

The efficiencies are factorized as
\begin{equation}
\varepsilon = \varepsilon^{\text{REC}}\varepsilon^{\text{SEL$/$REC}}\varepsilon^{\text{TRIG$/$SEL}},
\end{equation}
\noindent where $\varepsilon^{\text{REC}}$ is the reconstruction efficiency ($\sim0.4\%$ for \KFM decays) including acceptance effects,
$\varepsilon^{\text{SEL/REC}}$ is the efficiency of selecting the reconstructed signal candidates, including the BDT and PID
cuts ($\sim 38\% $ for \KFM decays) and $\varepsilon^{\text{TRIG/SEL}}$ is the trigger efficiency for events that would have been otherwise selected for offline analysis ($\sim20\%$ for \KFM decays).
The efficiencies are calculated using simulated events with corrections derived from \Kspipi decays in data. 
Tracking efficiencies are obtained from $J/\psi\rightarrow\mup\mun$ in data and \Kspipi decays using a tag-and-probe method~\cite{LHCb-DP-2013-002}. 

Similarly, PID efficiencies are obtained using calibration samples of low-momentum muons from $\jpsi\rightarrow\mup\mun$ decays collected at the same time as the data used for the analysis, and divided into six samples according to different data-taking conditions. 
The trigger efficiencies obtained in simulation are validated with misidentified 
$K^0\rightarrow\pipm\mump\nuornubar_\mu$
decays, as done in Ref.~\cite{LHCb:2012vqi}. 

The signal yield $N_{\KFM}$ is obtained from a maximum-likelihood fit to the four-muon invariant-mass distributions of the signal candidates in the two trigger categories.
The invariant-mass distributions of background events in the signal channel is assumed to be an exponential function. This assumption is validated using background events with lower values of the BDT output. The \KFM signal invariant-mass shape is parameterized using a Hypatia distribution~\cite{Hypatia}. The parameter values of the signal distribution are obtained from 
simulated events with corrections obtained from \Kspipi decays in data, following Ref.~\cite{LHCb:2012vqi}.

\section{Systematic uncertainties}
\label{sec:systematics}

The systematic uncertainties are incorporated
as Gaussian constraints to the fit of the signal-candidate mass distribution. For convenience, systematic sources are separated into those acting independently on each trigger category and those affecting the overall normalisation.
\tabref{tab:SystNorm} shows the main systematic uncertainties determined for this analysis.
The different sources of systematic uncertainties are summarized as follows:

\begin{itemize}
  \item The uncertainty on the branching fraction \BRof\Kspipi $ = (69.20 \pm 0.05) \%$ induces a $0.07\%$ uncertainty on the normalisation.
  \item The uncertainty on the average minimum bias prescale factor $s_{\text{MB}}$ is 0.3\%.
  
  \item A $1\%$ systematic accounts for small changes throughout the Run~2 data taking. These are mostly due to changes in the trigger settings, 
  and were determined by comparing the yields of misidentified triggered 
  $K^0\rightarrow\pipm\mump\nuornubar_\mu$ 
  decays with \Kspipi decays from the minimum bias data.
  \item To account for residual differences  between data and simulation, the efficiency ratio is recomputed with and without simulation corrections to the kaon momenta and track quality parameters. Half of the correction is taken as a systematic uncertainty ($4.4\%$).
  \item The systematic uncertainty on the the \Kspipi yield is evaluated by comparing the yield obtained from a fit to the $\pi^+\pi^-$ invariant mass with floating background components,
  with the default yield, which neglects the background contribution since the sample has very high purity. This effect is found to be less than $1\%$.
  \item The uncertainty is assigned as the difference between the maximum and minimum values of the six correction factors for the PID mentioned above.
  \item A systematic uncertainty associated to differences between data and simulation in the track reconstruction efficiency is determined using control channels from data.
  \item Potential differences in the trigger efficiencies between data and simulation are a source for systematic uncertainties. The L0 and HLT efficiencies for the \mbox{\KFM} signal are determined using misidentified 
  \mbox{$K^0\rightarrow\pipm\nuornubar$}decays as control modes. The efficiencies are found to be compatible between data and simulation, and the uncertainty in the validation is used as the systematic uncertainty, following Ref.~\cite{LHCb:2020ycd}. The  
  \mbox{$K^0\rightarrow\pipm\mump\nuornubar$} sample at LHCb is composed of \Ks and \Kl in similar amounts, due to a compensation between the different branching fractions~\cite{KLOE:2005vdt,KLOE:2005lau,KLOE-2:2019rev}, and the different reconstruction efficiency caused by the decay-time acceptance. The decay-time acceptance also results in similar decay-time distributions for the reconstructed \Kl and \Ks decays.
\end{itemize}

\begin{table}[!tp]
  \centering
  \caption{Summary of systematic uncertainties.}
  \label{tab:SystNorm}
  \begin{tabular}{l|c }
    Source                 & Relative effect $(\%)$                        \\
    \hline
    \BRof\Kspipi           & 0.07                                          \\
    $s_{\text{MB}}$                  & $0.30$                                \\
    Variations in data taking      & $1$                   \\
    Data/simulation differences & $4.40$                     \\
    \Kspipi yield    & $1$                                      \\
    PID                 & $3.30$    \\
    Tracking            &  $1.20$\\
    $\varepsilon^{\text{L0}}$     & $10$ (TIS), $21$ (xTOS)                      \\
    $\varepsilon^{\text{HLT/L0}}$ & $11$     \\
  
    \hline
    Total                  & $16$ (TIS), $24$ (xTOS) \\ 
  \end{tabular}
\end{table}

In addition, a systematic uncertainty of $1.4\%$ is added to the relative efficiency between \mbox{\KFM} and \mbox{\KlFM} decays, when measuring \mbox{\BRof\KlFM}. This systematic uncertainty covers the difference of the efficiency ratio across the data-taking period as well as the uncertainties from the simulated sample size. The systematic uncertainties from signal and background invariant-mass models are negligible for the current level of precision. 

\section{Results}
\label{sec:results}

The number of \Kspipi decays in the dataset corrected by the $s_{MB}$ prescale factor is $(3.96 \pm0.04) \times 10^{12}$. The efficiency ratio $\varepsilon_{\Kspipi}/\varepsilon_{\KFM}$ is $27.0\pm4.3$ ($21.6\pm5.3$) for the TIS (xTOS) category.
The obtained normalisation factors are \mbox{$\alpha_{\text{TIS}} = (4.7\pm 0.8) \times 10^{-12}$} and $\alpha_{\text{xTOS}} = (3.8\pm 0.9) \times 10^{-12}$. The combined single event sensitivity is $\alpha \approx 2.10\times 10^{-12}$. These values assume that the muons are distributed uniformly in phase space for \mbox{\KFM} decays. Nevertheless, the efficiency is nearly uniform as a function of the muon helicity angles and for most of the dimuon invariant masses, except near the kinematic thresholds. Thus, the results are valid for a wide range of BSM models unless the dimuons originate from a new resonance with an invariant mass close to half of the \Ks mass or very close to twice the muon mass. The former strengthens the limit while the latter weakens it. The result also assumes that the muons are produced at the $K^0$ decay vertex.

\begin{figure}
    \centering
    \includegraphics[width=0.96\textwidth]{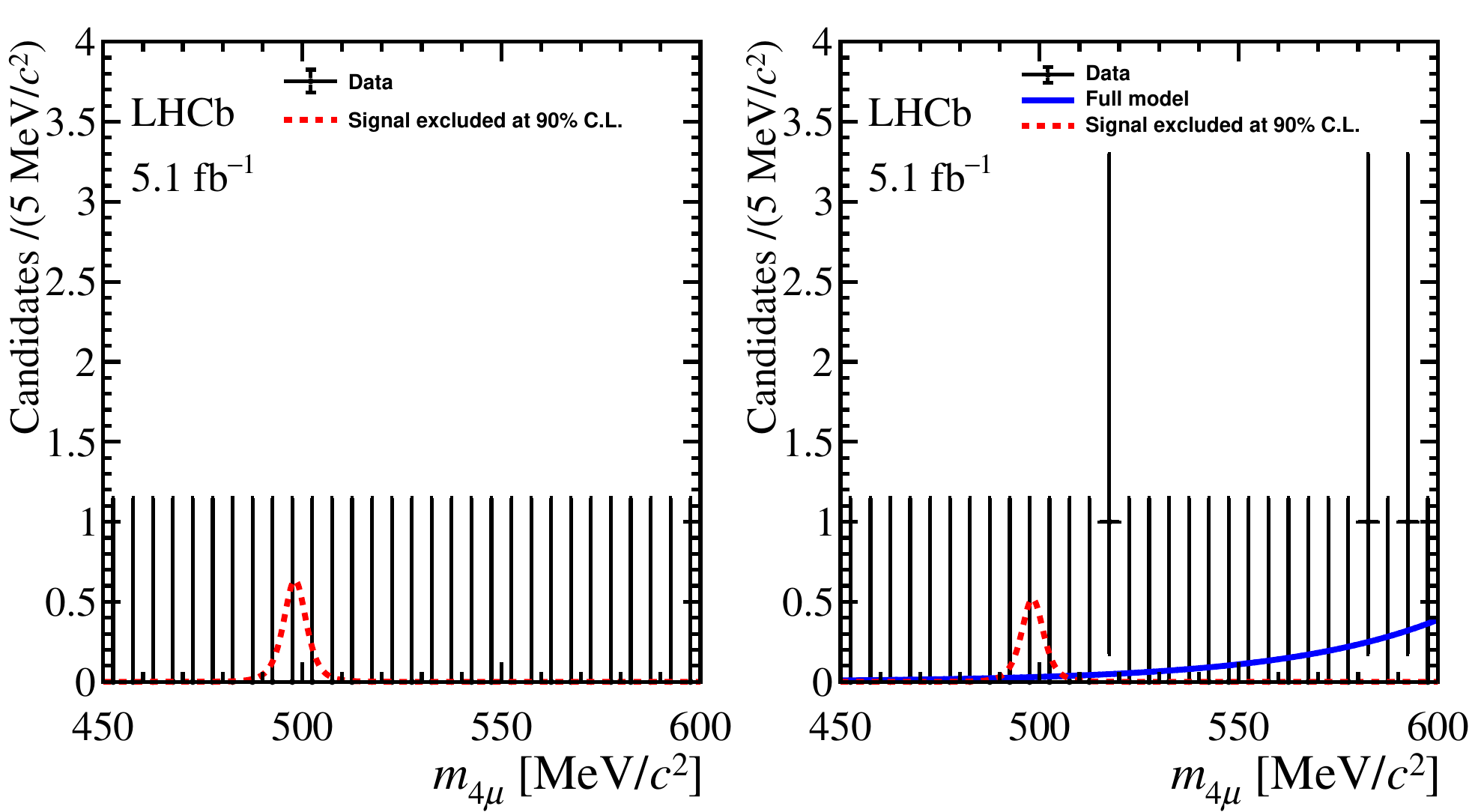}

    \caption{Invariant-mass distribution of the observed \KFM candidates in the (left) xTOS trigger category, and (right) TIS trigger category. The blue lines represent the simultaneous fit to both categories, using the exponential functions to represent the background. The expected \KFM signal for the branching fraction excluded at 90\% C.L. is shown with a dotted red line.}
    \label{fig:fit}
\end{figure}

The invariant-mass distributions of the selected candidate events are shown in \figref{fig:fit}. No events are observed in the signal mass window $490-510~\mevcc$, for both the xTOS and TIS samples. Integrating the profile likelihood from the positive side of the branching fraction, the upper limits at 90\% C.L. are found to be
\begin{align}
    \BRof\KFM &< 5.1 \times 10^{-12}, \nonumber \\
    \BRof\KlFM &< 2.3 \times 10^{-9} \nonumber , 
\end{align}
\noindent which coincide with the upper limits expected in absence of signal. 

While the data used to produce the results here were collected using a 1 MHz hardware trigger, the upgraded LHCb detector fully reconstructs events at the LHC inelastic event rate of 30 MHz~\cite{LHCb-TDR-021}. This will increase the trigger efficiency of the \KFM events by about a factor five~\cite{LHCb-TDR-021}, giving the possibility to reach sensitivities at the SM level, $\BRof\KFM\sim(1-4)\times10^{-14}$, with 300\invfb of integrated luminosity.

\section{Conclusions}
\label{sec:conclusions}

A search for \KzFM has been performed analysing $5.1\invfb$ of LHCb data recorded from 2016 to 2018. No signal is observed.
The obtained upper limits are the first reported for the \KzFM decay modes. The observed values are close to the maximum values allowed in the dark photon models~\cite{Goudzovski:2022vbt} and indicate very good prospects for the LHCb upgrade, which could achieve sensitivities at the level of the SM prediction for \BRof\KFM using 300\invfb of integrated luminosity.

\clearpage
\newpage












\section*{Acknowledgements}
%
%
\noindent We express our gratitude to our colleagues in the CERN
accelerator departments for the excellent performance of the LHC. We
thank the technical and administrative staff at the LHCb
institutes.
We acknowledge support from CERN and from the national agencies:
CAPES, CNPq, FAPERJ and FINEP (Brazil); 
MOST and NSFC (China); 
CNRS/IN2P3 (France); 
BMBF, DFG and MPG (Germany); 
INFN (Italy); 
NWO (Netherlands); 
MNiSW and NCN (Poland); 
MEN/IFA (Romania); 
MICINN (Spain); 
SNSF and SER (Switzerland); 
NASU (Ukraine); 
STFC (United Kingdom); 
DOE NP and NSF (USA).
We acknowledge the computing resources that are provided by CERN, IN2P3
(France), KIT and DESY (Germany), INFN (Italy), SURF (Netherlands),
PIC (Spain), GridPP (United Kingdom), 
CSCS (Switzerland), IFIN-HH (Romania), CBPF (Brazil),
Polish WLCG  (Poland) and NERSC (USA).
We are indebted to the communities behind the multiple open-source
software packages on which we depend.
Individual groups or members have received support from
ARC and ARDC (Australia);
Minciencias (Colombia);
AvH Foundation (Germany);
EPLANET, Marie Sk\l{}odowska-Curie Actions and ERC (European Union);
A*MIDEX, ANR, IPhU and Labex P2IO, and R\'{e}gion Auvergne-Rh\^{o}ne-Alpes (France);
Key Research Program of Frontier Sciences of CAS, CAS PIFI, CAS CCEPP, 
Fundamental Research Funds for the Central Universities, 
and Sci. \& Tech. Program of Guangzhou (China);
GVA, XuntaGal, GENCAT and Prog.~Atracci\'on Talento, CM (Spain);
SRC (Sweden);
the Leverhulme Trust, the Royal Society
 and UKRI (United Kingdom).

\clearpage
\newpage

\newpage


\addcontentsline{toc}{section}{References}
\bibliographystyle{LHCb}
\bibliography{main,LHCb-PAPER,LHCb-DP,LHCb-TDR,standard}

\newpage
\centerline
{\large\bf LHCb collaboration}
\begin
{flushleft}
\small
R.~Aaij$^{32}$\lhcborcid{0000-0003-0533-1952},
A.S.W.~Abdelmotteleb$^{50}$\lhcborcid{0000-0001-7905-0542},
C.~Abellan~Beteta$^{44}$,
F.~Abudin{\'e}n$^{50}$\lhcborcid{0000-0002-6737-3528},
T.~Ackernley$^{54}$\lhcborcid{0000-0002-5951-3498},
B.~Adeva$^{40}$\lhcborcid{0000-0001-9756-3712},
M.~Adinolfi$^{48}$\lhcborcid{0000-0002-1326-1264},
P.~Adlarson$^{77}$\lhcborcid{0000-0001-6280-3851},
H.~Afsharnia$^{9}$,
C.~Agapopoulou$^{13}$\lhcborcid{0000-0002-2368-0147},
C.A.~Aidala$^{78}$\lhcborcid{0000-0001-9540-4988}
% ...

{\footnotesize \it

% ...
$^{9}$Universit{\'e} Clermont Auvergne, CNRS/IN2P3, LPC, Clermont-Ferrand, France\\
% ...
$^{13}$LPNHE, Sorbonne Universit{\'e}, Paris Diderot Sorbonne Paris Cit{\'e}, CNRS/IN2P3, Paris, France\\
% ...
$^{32}$Nikhef National Institute for Subatomic Physics, Amsterdam, Netherlands\\
% ...
$^{40}$Instituto Galego de F{\'\i}sica de Altas Enerx{\'\i}as (IGFAE), Universidade de Santiago de Compostela, Santiago de Compostela, Spain\\
% ...
$^{44}$Physik-Institut, Universit{\"a}t Z{\"u}rich, Z{\"u}rich, Switzerland\\
% ...
$^{48}$H.H. Wills Physics Laboratory, University of Bristol, Bristol, United Kingdom\\
% ...
$^{50}$Department of Physics, University of Warwick, Coventry, United Kingdom\\
% ...
$^{53}$School of Physics and Astronomy, University of Glasgow, Glasgow, United Kingdom\\
$^{54}$Oliver Lodge Laboratory, University of Liverpool, Liverpool, United Kingdom\\
% ...
$^{62}$Syracuse University, Syracuse, NY, United States\\
% ...
$^{77}$Department of Physics and Astronomy, Uppsala University, Uppsala, Sweden, associated to $^{53}$\\
$^{78}$University of Michigan, Ann Arbor, MI, United States, associated to $^{62}$\\
\bigskip
% ...
}
\end{flushleft}

\end{document}